\documentclass[conference]{IEEEtran}
\usepackage[capposition=bottom]{floatrow}
\usepackage{cite}
\usepackage{amsmath,amssymb,amsfonts}
\usepackage{algorithmic}
\usepackage{graphicx}
\usepackage{textcomp}
\usepackage{xcolor}
\def\BibTeX{{\rm B\kern-.05em{\sc i\kern-.025em b}\kern-.08em
    T\kern-.1667em\lower.7ex\hbox{E}\kern-.125emX}}
\usepackage{tikz}
\usepackage{amsmath}
\usepackage{float}
\usepackage{pgfplots}
\pgfplotsset{compat=1.17} 
\usepackage{listings}
\usepackage{multirow}
\usepackage{url}
\usepackage{hyperref}
\usepackage{tabularx}
\usepackage{caption}
\usepackage{subcaption}
\begin{document}
\title{A Lightweight Routing Layer Using a Reliable Link-Layer Protocol}
\author{\IEEEauthorblockN{1\textsuperscript{st} Qiangfeng Shen}
\IEEEauthorblockA{\textit{Department of
Electrical \& Computer Engineering} \\
\textit{University of Toronto}\\
Toronto, Canada \\
qianfeng.shen@mail.utoronto.ca}
\and
\IEEEauthorblockN{2\textsuperscript{nd} Paul Chow}
\IEEEauthorblockA{\textit{Department of
Electrical \& Computer Engineering} \\
\textit{University of Toronto}\\
Toronto, Canada \\
pc@eecg.toronto.edu}
}



\maketitle

\begin{abstract}
In today's data centers, the performance of interconnects plays a pivotal role. However, many of the underlying technologies for these interconnects have a history of several decades and existed long before data centers came into being.To better cater to the requirements of data center networks, particularly in the context of intra-rack communication, we have developed a new interconnect. This interconnect is based on a lossless link layer protocol, named RIFL. In this work, we designed and implemented RIFL Layer 2, a scalable network that supports up to multi-hundred Gbps communication. RIFL Layer 2 includes the RIFL switch and RIFL NIC. By utilizing a simple Batcher Banyan and iSLIP RIFL switch, we effectively keep the typical intra-rack latency under 400 nanoseconds. Moreover, for a 32-port 100Gbps network, under both Bernoulli arrival and bursty arrival traffic patterns, we ensure that the 99\% tail latency does not exceed 12microseconds.
\end{abstract}

\begin{IEEEkeywords}
Network switch, Low latency, Lossless switching
\end{IEEEkeywords}

%
\IEEEpeerreviewmaketitle

\section{Introduction}

In recent years, because of the advancements in cloud computing, artificial intelligence, and other data-intensive applications and services, there is an unprecedented surge in the volume and complexity of the data center workloads. As a result, data centers are faced with growing challenges related to performance and scalability. In this context, the communication infrastructure within a data center plays a critical role in ensuring efficient data movement and resource sharing among interconnected components.

One particular area that demands attention is intra-rack communication—the data transmission and exchange that occurs within a rack, encompassing servers, storage devices, accelerators, and other computing resources residing within a physical enclosure. Studies\cite{benson2010network} have shown that, on average, 75\% of the data center traffic is confined within the rack. Traditionally, intra-rack communication has relied on established protocols such as Ethernet~\cite{metcalfe1976ethernet} or InfiniBand~\cite{Infiniband}. However, the rapid evolution of data center workloads and the emergence of new technologies have exposed limitations in the existing protocol suites, prompting the need for a new set of protocols specifically tailored to intra-rack communication.

Many of the traditional network protocol stacks used in data centers today were invented decades ago, before data centers were existed. Some of these protocols, such as TCP(UDP)/IP, are not only used for data center networking but also for reliably delivering data from one corner of the world to another. As a result, these protocols may not reflect the needs of today's data centers. 

Although all protocols have to sacrifice some performance to ensure a certain level of orderly, reliable transmission, the cost to achieve this in an intra-rack communication context is completely different. There are two main factors affecting the reliability of transmission: one is the noise in the physical channel between network nodes causing bit errors and thus errors in the transmitted data; the other is the overflow of switch buffer queues due to network congestion. Both TCP/IP and InfiniBand choose to use end-to-end retransmission to address the first issue. 

When using end-to-end retransmission, inevitably, each data receiver needs to continuously send acknowledgements (ACKs) to the data sender as long as data transmission does not stop. This leads to two drawbacks: first, acknowledgements will occupy the bandwidth of the reverse channel. Worse, during network congestion, if acknowledgements are dropped or do not reach the sender within the specified time, the sender will resend the content that the receiver has already received. This will further exacerbate network congestion, forming a positive feedback loop. Certainly, end-to-end retransmission also offers advantages. One such advantage is that, during network congestion, switches are not required to entirely prevent buffer overflow. Any data loss resulting from buffer overflow can be recovered using end-to-end retransmission, although this comes with the trade-off of reduced network performance.

Moving retransmission from the higher-level protocol stack to the link layer and replacing end-to-end retransmission with segmented (hop-by-hop) retransmission is an approach that may significantly improve overall network performance. Compared to end-to-end retransmission, segmented retransmission can rely solely on Negative Acknowledgements (NAKs) without occupying the bandwidth of the reverse channel when no bit errors occur. Moreover, since retransmission is limited to the two nodes connected by a network cable, it does not affect the congestion of the switches. RIFL~\cite{shen2022rifl}~\cite{shen2021rifl} is one such link layer protocol that follows this principle. RIFL achieves lossless data transmission with significantly lower latency and bandwidth overhead than Forward Error Correction (FEC). The core concept of RIFL bears similarities to the FLIT mechanism in PCIe 6.0~\cite{pcie6}, but with the added advantage of eliminating the need for ACKs.

Expanding upon the groundwork laid by RIFL, this study presents the design of RIFL Layer 2—a lightweight routing layer, the RIFL switch—which features a Batcher-Banyan switch combined with an iSLIP~\cite{mckeown1999islip} scheduler, and an implementation of a Data Plane Development Kit~\cite{dpdk} (DPDK)-style RIFL Linux driver. Although our long-term plan is to apply the RIFL network across the entire data center network, in this work, our attention is particularly directed towards intra-rack network communication. As a result of this work, high-performance lossless data transmission based on RIFL for intra-rack communication becomes possible. Note that although used in intra-rack communication, the positioning of the RIFL network is different from protocols such as Compute Express Link~\cite{van2019hoti} (CXL). CXL operates based on PCIe, focusing on high level functionalities such as providing efficient remote memory access semantics. In contrast, the RIFL network is more dedicated to providing low-latency, reliable communication at a lower level.

The rest of this paper is organized as follows: Section \ref{SECTION:Background} briefly introduces RIFL and the conventional switch architectures. In Section \ref{SECTION:relatedwork} we discusses the related works. In Section \ref{SECTION:L2}, we define the RIFL Layer 2 and present our switch architecture. Section \ref{SECTION:results} provides performance results. Section \ref{SECTION:conclusion} concludes this work, and Section \ref{SECTION:futurework} discusses our upcoming work.\section{Background}
\label{SECTION:Background}

In this section, we will first introduce how RIFL works and how it achieves lossless transmission with extremely low overhead. Then, we will introduce a series of classic switch architectures and explain why we chose the Batcher-Banyan network combined with iSLIP scheduler to implement a RIFL switch.

\subsection{RIFL}
As previously mentioned, RIFL is a link layer protocol designed for lossless data transmission. It can be used in point-to-point links with multi-hundred Gbps throughput. At 100 Gbps, it features a latency of approximately 40 ns in addition to the delay introduced by transceivers and network cables. It segments the data being transmitted into frames of fixed sizes. While the frame size is fixed, RIFL allows for different implementations to opt for varying frame sizes, ranging from 8 bytes to 256 bytes, provided the frame size is a power of two. Choosing larger frame sizes reduces bandwidth overhead but increases latency. Each RIFL frame includes a two-byte header, which is independent from the frame size of the implementation. Using two encoding schemes - "meta code" and "verification code" - this small header carries five pieces of information: 1. the sync word for the line code, 2. the number of valid bytes within the frame, 3. the end-of-packet (EOP) indicator, 4. the checksum, which is a twelve-bit cyclic redundancy code (CRC), and 5. the frame sequence number. With this information, RIFL can reliably transmit data across the link at byte level granularity.

\begin{figure*}[hbt!]
  \centering
  \begin{tikzpicture}
    \node (fig) {
      \includegraphics[width=0.9\textwidth]{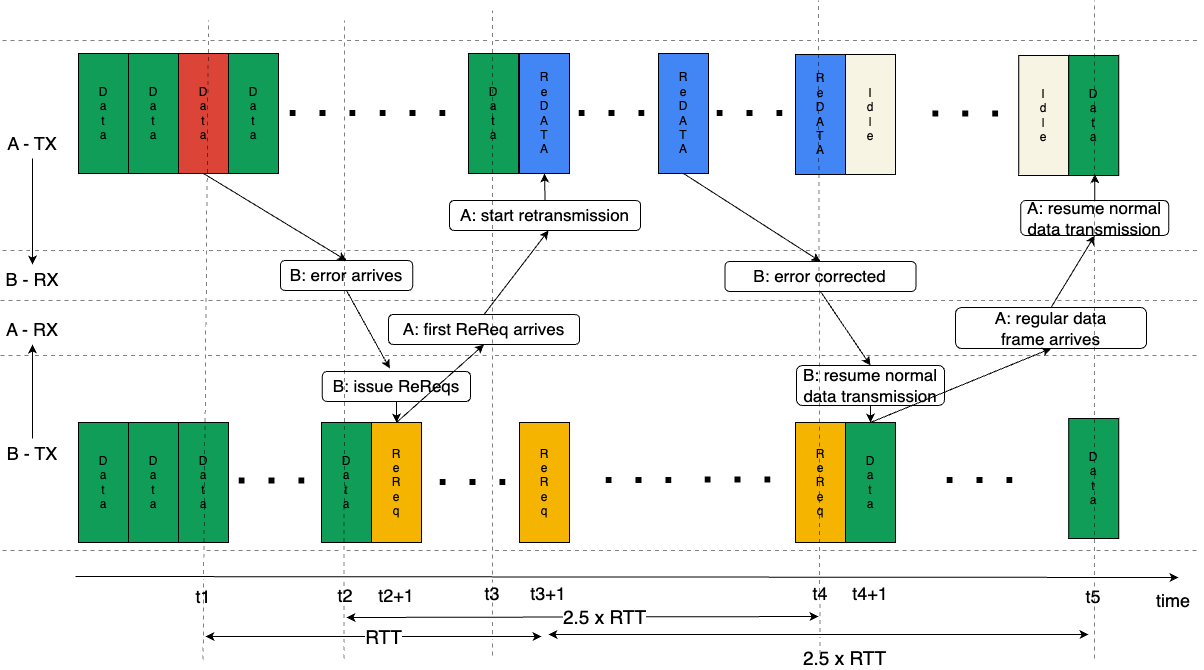}
    };
    \node (A) at (fig.south west) [anchor=north west,color=darkgray, minimum width=0.5\textwidth] {\footnotesize
        Data in green: Data frame with no error \space\space\space\space\space\space\space\space\space\space\space
        ReReq: retransmission request frame
        \space\space\space\space\space\space\space\space\space\space\space
        Idle: idle frame
    };
    \node (B) at (A.south west) [anchor=north west,color=darkgray] {\footnotesize
        Data in red: Data frame with error
        space\space\space\space\space\space\space\space\space\space\space\space
        ReData: retransmitted data frame
    };
  \end{tikzpicture}
  \caption{\label{fig:rifl}Schematic Diagram of RIFL Retransmission}
\end{figure*}

The core idea of RIFL is simple: the bit error rate (BER) for a single link is relatively low (for example, IEEE 802.3~\cite{IEEE802.3} requires that the BER for 100 Gbps Ethernet does not exceed $10^{-12}$). Moreover, within the physical dimensions of a data center, the transmission delay between two directly connected nodes can generally be maintained within a range of a few hundred nanoseconds. With such a low BER and point-to-point latency, even for links with line rates exceeding 100 Gbps, it is possible to pause bidirectional communication when errors occur, spend one to two microseconds retransmitting the erroneous frames, and then resume normal data transmission without affecting the link's bandwidth efficiency, typical latency, and tail latency. RIFL~\cite{shen2022rifl} research demonstrates that for a 100 Gbps link with a distance of up to 500 meters between the two nodes, as long as the BER is lower than $10^{-7}$, RIFL can ensure that the data transmission performance is not less than 99\% of the performance when the BER is 0.

RIFL achieves this by employing a novel retransmission scheme. As shown in Figure~\ref{fig:rifl}, a complete RIFL retransmission process can be described as follows, assuming device A and device B are connected by a RIFL link:

\begin{enumerate}
    \item At time point t1, A transmits a data frame to B. During the transmission process, this data frame experiences errors, and it arrives at B at time point t2, half a round trip time (RTT) later.
    \item At time point t2, B detects the errors and stops sending normal data frames in the next cycle, instead continuously sending retransmission requests to A. The first retransmission request sent by B will arrive at A at time point t3, half an RTT later.
    \item At time point t3, A detects the errors and stops sending normal data frames in the next cycle, instead starting to retransmit in sequence. This process takes 2.5 RTTs. One RTT will be used to resend the data frames sent from A to B within the previous RTT before t3. Another 1.5 RTTs will be used to send control information. This control information is used to ensure that even if errors occur simultaneously in bidirectional transmission, retransmissions can proceed in both directions. 
    \item At time point t4, the previously erroneous data frame is retransmitted to B, and B will resume sending regular data frames in the next cycle.
    \item At time point t5, A completes a 2.5 RTT retransmission cycle. If A no longer receives retransmission requests sent by B at this point, B will stop retransmitting and resume sending normal data frames.
\end{enumerate}

To correct an error, both nodes will pause the transmission of normal data for approximately 2.5 round trip times (RTT). The process from the occurrence of an error to its correction takes about 3.5 RTTs.

In addition to retransmission, RIFL also employs a simple ON/OFF flow control to allow the receiver to deliver back pressure to the sender. The flow control commands are protected by the checksum to avoid being missed. This ensures that links using RIFL for transmission remain error-free and without packet loss. It is worth noting that the original flow control mechanism in RIFL would pause all flows being transmitted on the link. To provide better granularity, we extended the original flow control mechanism: we configured a virtual output queue (VOQ) for each input port of the RIFL switch. Our new flow control mechanism allows the switch port to send flow control information for each individual virtual queue to the other end. We will introduce the detailed mechanism in Section~\ref{SECTION:L2perflowcontrol}.

In summary, RIFL is a lossless link layer protocol incorporating extremely low latency. Protocols running on top of RIFL no longer need to carry checksums within their packet headers. The simplest protocol running on a RIFL link could be a FIFO.  Data pushed in at the input will appear at the output without error.  We believe that applying RIFL to intra-rack communication will potentially result in significant performance improvements.

\subsection{Switch Architectures}
The architecture of switches has undergone several generations of development. To achieve better scalability and support for more ports, many modern switch designs~\cite{scott2006blackwidow,dong2011non,chrysos2006scheduling,sapountzis2005benes,oki2002concurrent,chaney1997design} choose multi-stage network architectures (e.g., a Clos network) instead of a crossbar. This is because, compared to crossbar structures, multi-stage network architectures can reduce the number of cross points from $N^2$ to $N^{1.5}$ (or even to $NlogN$ for the Benes network) while maintaining non-blocking properties. This not only allows the switch chip to maintain a reasonable size but also alleviates the pressure on place and route.

Our goal is to support multi-hundred Gbps interconnects, which naturally leads us to choose a multi-stage switch architecture. However, our aim is not to select the best-performing switch architecture in the traditional sense, as the highest-performing architecture may not necessarily be the most suitable for the RIFL and intra-rack communication context. What we need is a non-blocking switch that maintains RIFL's low latency, has low complexity, is lightweight, and is capable of providing good throughput within the rack.

With the aforementioned properties in mind, we attempted to analyze some established multi-stage switch architectures. For multi-stage switches, there are typically three major stages: the input stage, the central stage, and the output stage. As the names suggest, switches belonging to the input stage are adjacent to the ingress ports, switches belonging to the output stage are adjacent to the egress ports, and switches belonging to the central stage are located in between the two. 

Based on whether each stage has a buffer or not, different multi-stage switch architectures can be divided into eight distinct types, ranging from Space-Space-Space (SSS) to Memory-Memory-Memory (MMM). "Space" means that the corresponding stage has no buffer, while "Memory" means that the corresponding stage has buffers. Among them, MMM, MSM, and MSS are more popular. MSM and MSS usually require complex scheduling algorithms to achieve better performance. MMM can typically reach higher performance, however it requires a large amount of buffers and may introduce problems such as out-of-sequence packet delivery. For example, Oki et al.~\cite{oki2002concurrent}'s MSM switch requires multi-stage schedulers between each pair of input switches and central switches; As for MMM, Sapountzis et al.~\cite{sapountzis2005benes} require equipping each switch in the Benes network with a VOQ, while Chrysos~\cite{chrysos2006scheduling} et al. assume that each VOQ in the ingress line card is implemented using DDR SDRAM. Moreover, under light traffic Chrysos et al.'s approach cannot provide low latency. It is evident that these switch architectures are either very complex or unable to guarantee low latency. 

We decided to adopt a simpler architecture: We know that the iSLIP~\cite{mckeown1999islip} scheduler performs well on a crossbar switch. 
Although we cannot use a crossbar switch due to potential routing difficulties at speeds of hundreds of Gbps, we can utilize a Batcher-Banyan network~\cite{narasimha1988batcher}, which is functionally equivalent to a crossbar if and only if at each time slot the destination port of the traffic for each input port is unique. The iSLIP scheduler is able to ensure this condition is met, but it faces challenges scaling up for switches with more than 64 ports. However, typically, 64 ports are sufficient for a single rack~\cite{portCountEffect}. 

In summary, we choose to use the Batcher-Banyan network combined with the iSLIP scheduler as our switch architecture due to considerations of complexity, cost, performance, and the application environment.\section{Related Work}
\label{SECTION:relatedwork}
At the current stage, we position RIFL network as an intra-rack communication interconnect. It can transmit data losslessly between endpoints with extremely low latency. In this section, we list three related works that are functionally similar to the RIFL network. However, at the same time, they have fundamental and significant differences from the RIFL network. This is because the RIFL network's unique and efficient data link layer retransmission protocol allows its upper layers to be simpler, and the buffer queues in RIFL switchs can be shallower.

\subsection{Peripheral Component Interconnect Express (PCIe)}
PCIe~\cite{pcie6} has a history spanning several decades. As the successor to PCI, PCIe is more similar to RIFL network as it is also based on point-to-point links, and its expansion is achieved through the PCIe root complex and PCIe switch. However, its scalability is often confined within a backplane. PCIe links are not designed to be peer-to-peer. They are categorized into upstream and downstream links. The CPU is always the central element positioned upstream. Although downstream peripheral devices can communicate with each other over PCIe, they must first be enumerated by the CPU. The latest PCIe standard includes a link layer retransmission mechanism named FLIT. However, it differs from RIFL in that it uses acknowledgements, while RIFL employs negative acknowledgements. In the RIFL paper~\cite{shen2022rifl}, it is discussed that using negative acknowledgements can potentially achieve higher bandwidth efficiency and lower latency.

\subsection{InfiniBand}
InfiniBand~\cite{Infiniband}, as a high-speed interconnect technology, utilizes a reliable transport protocol called Reliable Datagram Sockets (RDS) to handle retransmissions.

RDS uses acknowledgments (ACKs) to ensure reliable delivery of messages. When a sender transmits a message, it waits for an ACK from the receiver. If the sender does not receive an ACK within a specified timeout period, it assumes that the message was lost or not delivered. In the case of a missing message, the sender retransmits the message to ensure its delivery. The sender keeps track of sent messages and maintains a retransmission queue for pending ACKs. When the receiver detects a missing message, it sends a Negative ACK (NACK) to the sender, requesting retransmission of the missing message. Upon receiving a NACK, the sender retransmits the missing message from its retransmission queue. It continues to retransmit until it receives a positive ACK or reaches a maximum retry limit. The receiver, upon receiving duplicate messages, discards redundant duplicates based on message sequence numbers. The retransmission process continues until all missing messages are successfully delivered and acknowledged.

RDS differs from RIFL's retransmission and PCIe's FLIT as it handles retransmission at the transaction layer. As discussed in ~\cite{shen2022rifl}, in a data center environment, link layer retransmission has significant advantages over transaction layer retransmission.
\subsection{A customized Ethernet based protocol}
Sanchez et al. \cite{micro} build an interconnect for FPGA-based high performance computing. Their protocol is based on the 10 Gigabit Media Independent Interface (XGMII). The throughput is limited as 10 Gbps. Switching and Routing in their network is performed by a Ethernet switch. This work is based on the assumption that the link channels are error free, therefore it is not a lossless network by itself.\section{RIFL Layer 2}
\label{SECTION:L2}
In this section, we will introduce the second layer protocol of RIFL (hereinafter referred to as Layer 2), the custom-designed PCIe-DMA engine and Linux driver that facilitate CPU integration with RIFL, the implementation of a RIFL switch, and the modification we made to the flow control mechanism of the RIFL link layer to achieve flow control for each virtual output queue.
\subsection{The Layer 2 Protocol}
\label{SECTION:L2protocol}
\begin{figure}
\includegraphics[width=\textwidth]{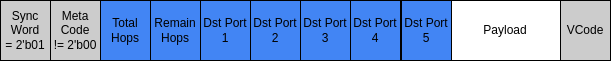}
\caption{Packet format for RIFL Layer 1 and 2}
\label{FIG:L2Header}
\end{figure}

\begin{figure*}[hbt!]
\includegraphics[width=0.9\textwidth]{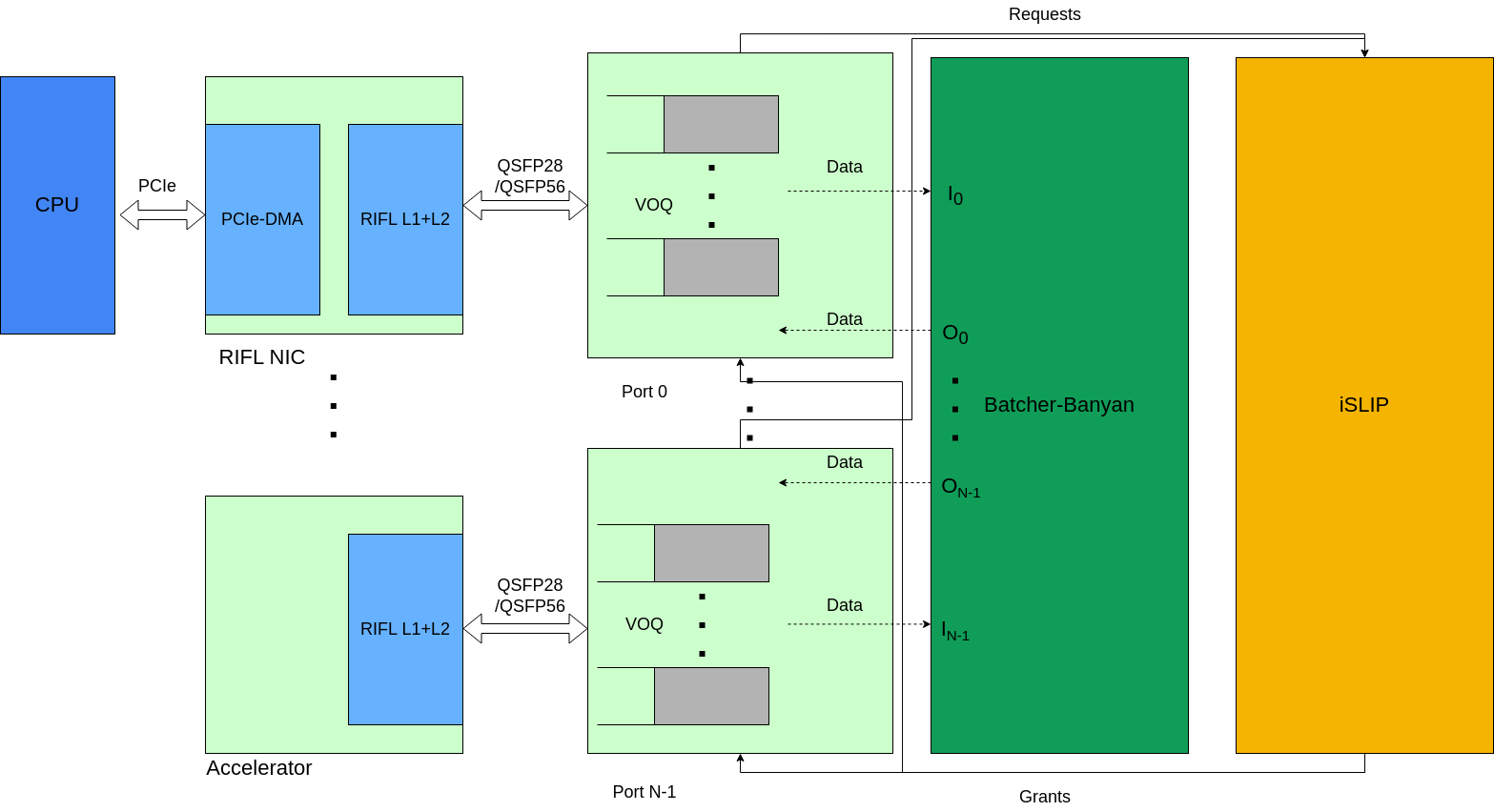}
\caption{RIFL Layer 2 switch}
\label{FIG:L2switch}
\end{figure*}

To ensure that Layer 2 is compact, we decided that its header will only contain information used for routing. Any additional information can be reflected in the higher-level protocols running on top of it.

After an upper layer packet is delivered to Layer 2, it is divided into 256-byte segments, each of the segments is a Layer 2 packet. If the size of the upper layer packet is not a multiple of 256 bytes, the last segment will be padded to 256 bytes. There are two benefits to this approach. First, switches that only support forwarding fixed-size packets are much simpler and more efficient than those that support forwarding variable-length packets~\cite{cisco}. Second, due to the characteristics of RIFL Layer 1, if a switch port detects a bit error, it will not forward any valid data for the next one to two microseconds. If the port had previously been competing with other ports for the same output port and had won arbitration, it means that during the process of resolving the bit error, it will prevent other ports from forwarding data to the corresponding output port, and the output port will not receive any data during this process. By using fixed and small size packets, we can have each input port participate in arbitration after receiving a complete packet, thus avoiding the blocking mentioned previously with a small latency added to each hop. The reason for choosing 256-byte segments instead of smaller ones is that it allows the iSLIP scheduler more time to complete more iterations, leading to better performance. Moreover, unlike other switches that choose to implement segmentation and de-segmentation within the switch, we decided to delegate these two steps to the endpoints. This approach not only simplifies the switch, but also results in shorter latency, since segmentation and de-segmentation only need to be done once.

Figure~\ref{FIG:L2Header} shows the packet format of Layer 2. The gray fields are the header fields of RIFL Layer 1, and the blue fields are the header fields of Layer 2. The functions of each Layer 2 header field are as follows: 
\begin{itemize}
    \item Total Hops: the total number of hops for this packet. This field consists of four bits. 
    \item Remain Hops: the remaining number of hops for this packet. This field also consists of four bits.
    \item Dst Port i ($i~\epsilon~[1,5]$): routing information for the corresponding hop. There are five such fields in a Layer 2 header, each field consists of eight bits.
\end{itemize}

The headers of RIFL Layer 1 and Layer 2 combined are a total of eight bytes. Unlike traditional protocols such as Ethernet, Layer 2 no longer assigns a fixed physical address to each endpoint. Instead, communication between two endpoints uses relative addresses. The Layer 2 header shown in Figure~\ref{FIG:L2Header} allows a packet to travel up to five hops, which we believe is enough for intra-rack communication. In each hop, limited by the size of the ``DST PORT" fields, the switch can have up to 256 ports. When each packet is sent out by the endpoint, its ``Total Hops" and ``Remain Hops" will be written with the same value: the number of hops the packet is about to travel. When a packet enters the switch, the switch first checks if its “remaining hops” is equal to 0. If not, the packet needs to be forwarded to the corresponding destination port. The switch will examine the value of ``Dst Port 1" to determine the forwarding destination. The specific rule can be described by Equation~\ref{eq:dport}. Assuming in an N-port switch, the value of ``Dst Port 1" for the packet currently being forwarded by input port $i$ is $j$, the destination port $d(i, j)$ to which the packet should be forwarded is:
\begin{equation}\label{eq:dport}
d(i,j) = 
    \begin{cases}
        j, & \text{if } j < i\\
        j+1, & \text{if } i \leq j < N\\
        broadcast, & \text{if } j == 255\\
    \end{cases}, 
\end{equation}
Note that in RIFL switches, packets entering from an input port are not permitted to be looped back to the same port.

As each packet leaves the input port and is about to enter the switch fabric, its ``Remaining Hops" will be decremented by one, and all ``Dst Port" fields will be shifted one byte to the left. ``Dst Port 2, 3, 4" will be moved to the positions of ``Dst Port 1, 2, 3", respectively, while ``Dst Port 1" will be moved to the position of ``Dst Port 4". When the packet arrives at the designated endpoint, by reversing the ``Dst Port" fields and examining the value of ``Total Hops", the endpoint can easily obtain the relative address of the packet's source. The advantage of using relative addresses is that during each switch forwarding process, it eliminates the need to map fixed physical addresses to the destination port. This step often involves Content Addressable Memory (CAM), which is an expensive resource.

Similar to how Internet Protocol (IP) functions with Ethernet, the upper layer protocols of Layer 2 might want to assign fixed logical addresses to different endpoints. These logical addresses can be mapped to Layer 2's physical addresses using a method similar to the Address Resolution Protocol~\cite{plummer1982arp} (ARP). Alternatively, a centralized address mapping server can be employed to send the mapping information to each endpoint through side-band channels, such as using the control plane.

In summary, in this subsection, we introduced the RIFL Layer 2 protocol, which encapsulates forwarding data into fixed-size packets and uses relative addresses to guide switch forwarding.

\subsection{The RIFL NIC}
\label{SECTION:L2Driver}

To enable CPUs to communicate with other CPUs and accelerators using RIFL, we have built a prototype of a RIFL NIC using an FPGA. In addition to containing the endpoint logic for RIFL Layer 1 and RIFL Layer 2, the hardware part of the RIFL NIC also includes a PCIe-DMA engine. This engine can convert DMA transactions into PCIe Transaction Layer Packets (TLPs). We connected it to AMD/Xilinx's ``PCIE4" IP core~\cite{PCIE4}, enabling non-blocking peer-to-peer DMA transfers between the RIFL NIC and the CPU. In theory, our DMA engine can work with any PCIe PHY after interface adaption. The software part of the RIFL NIC includes a set of DPDK-style user-space Linux drivers. These drivers utilize pre-allocated huge pages and PCIe Virtual Function I/O (VFIO), but do not rely on any DPDK libraries. On the FPGA prototype, the bandwidth between the CPU and the RIFL NIC can reach up to approximately 90 Gbps, with a minimal round trip delay of around $2.5 \mu s$.
\subsection{The RIFL Switch}
\label{SECTION:L2Switch}

The RIFL switch architecture is presented in Figure~\ref{FIG:L2switch}. The RIFL Switch mainly consists of three modules: Input/Output Modules, the iSLIP scheduler, and the Batcher-Banyan network. Among them, the Input/Output Modules can be considered as line cards, which primarily contain VOQs (Virtual Output Queues). To reduce complexity, each VOQ is implemented by a statically allocated multi-queue~\cite{tamir1988high}(SAMQ). The iSLIP scheduler is mainly responsible for scheduling. The iSLIP scheduler receives $N^2$ bits of request signals each time and outputs $N*log_2N$ bits of arbitration results within the specified time. The specific duration depends on the line rate. For example, for a 100 Gbps interconnect, since the size of each Layer 2 packet is 256 bytes, the iSLIP scheduler needs to make a decision within approximately 20 nanoseconds (packet size divide by line rate) after each time the requests are sent. For a 200 Gbps interconnect, this constraint tightens to 10 nanoseconds. Utilizing the implementation proposed by Gupta et al.~\cite{gupta1999designing}, we successfully implemented a 32-port iSLIP scheduler at 400MHz and a 64-port scheduler at 200MHz on a 7nm Xilinx Versal FPGA. This allows us to execute three iterations of iSLIP for the 32-port design and one iteration for the 64-port design. We believe this clock frequency could be significantly increased if we implement this module using ASICs rather than FPGAs. The Batcher-Banyan network is responsible for routing packets to the specified output port.

\begin{figure}
\includegraphics[width=\textwidth]{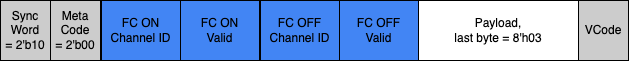}
\caption{Frame format for per-VOQ flow control}
\label{FIG:L2fc}
\end{figure}

\subsection{Per-VOQ Flow Control}
\label{SECTION:L2perflowcontrol}

Figure~\ref{FIG:L2fc} shows a modified RIFL Layer 1 flow control frame for per-VOQ flow control. Each such flow control frame can pause the traffic sent from an upstream device to at most one VOQ downstream while also being able to unpause traffic sending to another VOQ. Whenever a packet is stored in a VOQ from the switch's input port, if the data storage of that queue just exceeds the flow control ON threshold, the switch will send a flow control frame to the endpoint at the next moment. 
The ``FC ON Channel ID" in this frame will be the ID of the queue, and ``FC ON Valid" will be high. Similarly, when a packet leaves the VOQ, if the data storage of that queue just drops to the flow control OFF threshold, then at the next moment, the switch will also send a flow control frame to the endpoint. The ``FC OFF Channel ID" in this frame will be the ID of the queue, and ``FC OFF Valid" will be high. Because at any moment, at most one packet enters the VOQ and at most one packet leaves the VOQ, this flow control mechanism can precisely pause and unpause the data transmission for all VOQs.

The size of each queue here must also satisfy the conditions discussed in RIFL Layer 1~\cite{shen2022rifl}, that is, each virtual queue must be able to store all the data sent within at least 1.5 RTT. For a 100Gbps switch with 64 ports, assuming the RTT is 300 nanoseconds, the size of the VOQ required for each input port is about 350KB, and 64 ports need a total of about 22MB of memory, which is relatively reasonable.\section{Performance of the RIFL Network}
\label{SECTION:results}
In this section, we will present the performance results of the RIFL network. We have implemented the RTL design for the RIFL network. Our performance tests are based on the simulation results of the synthesizable 32-port 100Gbps RIFL switch design. 
We have also prototyped a 32-port 10G RIFL switch on Arista's 7130 FPGA-enabled switches~\cite{L1switch}, and a 4-port 100G RIFL switch on the AMD/Xilinx VCU128 platform~\cite{VCU128}.

\subsection{Simulation setup}
Since the RIFL switch is based on the iSLIP scheduler, to a large extent, the RIFL switch also inherits the performance characteristics of iSLIP. In the RTL level simulation, we simulated two traffic patterns: uniform Bernoulli arrival and bursty arrival. For both traffic patterns, we tested packets of fixed sizes as well as packets of variable sizes. The size of the fixed-size packets is a multiple of 256 bytes, while the size of the variable packets varies from 64 bytes to 2048 bytes. We modeled the traffic pattern using the same approach as described in the iSLIP~\cite{mckeown1999islip} paper.  Specifically, under the bursty traffic pattern, each bursty period contains a geometrically distributed number of cells, and in each burst period the expected amount of data to be transmitted over a single port is 1024 bytes. The length of the idle period also follows a geometric distribution, and it varies depending on the workload. For both bandwidth and latency tests, we configured each device connected to the input port to send 2.5 Megabytes (MBs) of data to every device connected to the output port - each input port and output port of the RIFL switch will transmit 80~MBs of data. 
The reason we set the data size for each flow at 2.5~MB is because we observed that, after the test data volume increases beyond this value, there is no longer a visible change in the network performance.

During the bandwidth tests, we measure the switch's aggregated bandwidth utilization under different workloads, which is the aggregated bandwidth over all ports divided by the total line rate of all ports. During the latency testing, we observe the distribution of latency across all packets. We set the latency of components other than the VOQ and the iSLIP scheduler to be 17 cell times. based on observations from real hardware tests. This latency includes the delay from the RIFL Layer 1, Ethernet cable, transceiver, etc. In RIFL layer 2, each cell is 256 Bytes, for 100 Gbps network, each cell time is 20 nanoseconds; for 200 Gbps network, each cell time is 10 nanoseconds.

In both the bandwidth and latency tests, we use an ideal statically allocated, fully connected (SAFC)~\cite{tamir1988high} switch as a comparison. An N-port SAFC switch has $N^2$ physical queues, ensuring a dedicated queue for each pair of input and output ports. Owing to its limited scalability, bringing an 32-port 100 Gbps SAFC switch to real life is deemed unfeasible. However, in this paper, we utilize its performance as a benchmark for the theoretical maximum performance a switch could achieve.

\subsubsection{Bandwidth}

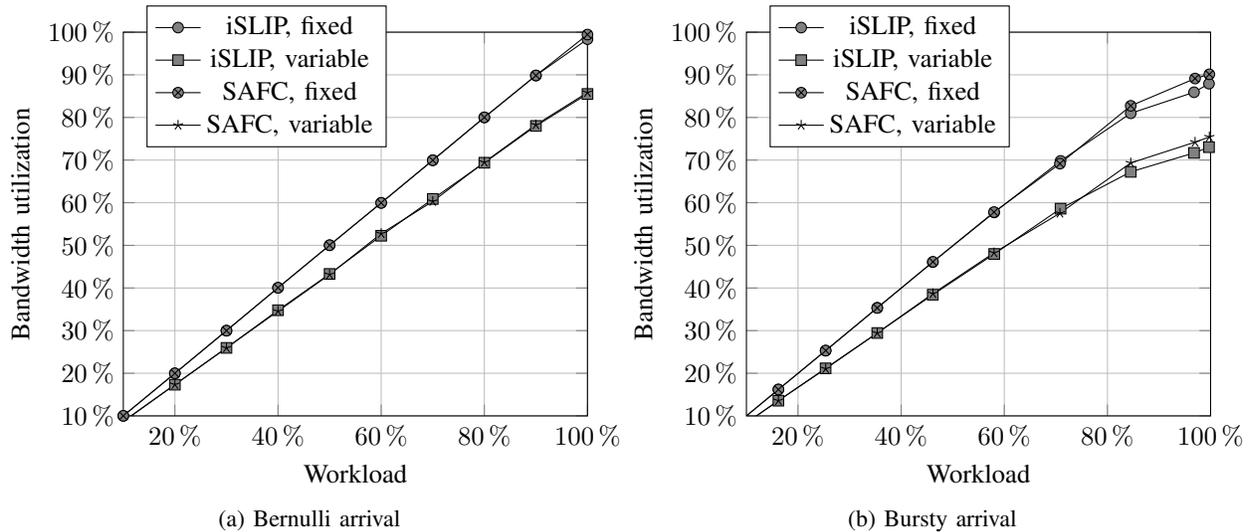
\begin{figure*}
\centering
\begin{subfigure}[b]{0.45\textwidth}
\begin{tikzpicture}
\begin{axis}[
    width=0.95\textwidth,
    legend style={at={(0.05,0.70)},anchor=south west},
    cycle list name=black white,
    xlabel={Workload},
    ylabel={Bandwidth utilization},
    xmajorgrids,
    ymajorgrids,
    zmajorgrids,
    xminorticks=true,
    xticklabel=\pgfmathprintnumber{\tick}\,$\%$,  
    yminorticks=true,
    yticklabel=\pgfmathprintnumber{\tick}\,$\%$,
    xmin=10, xmax=100,
    ymin=10, ymax=100,
    xtick={0,20,...,100},
    ytick={0,10,20,...,100},
]

\addplot coordinates {(10,10.01)(20,19.98)(30,29.99)(40,40.05)(50,50.03)(60,59.96)(70,69.96)(80,80)(90,89.83)(100,98.36)};
\addlegendentry{iSLIP, fixed}

\addplot coordinates {(10,8.639)(20,17.32)(30,25.95)(40,34.78)(50,43.29)(60,52.25)(70,60.88)(80,69.38)(90,78.01)(100,85.5)};
\addlegendentry{iSLIP, variable}

\addplot coordinates {(10,10.01)(20,19.98)(30,29.99)(40,40.05)(50,50.03)(60,59.96)(70,69.96)(80,80.02)(90,89.84)(100,99.44)};
\addlegendentry{SAFC, fixed}

\addplot coordinates {(10,8.731)(20,17.35)(30,25.95)(40,34.56)(50,43.16)(60,52.76)(70,60.31)(80,69.49)(90,78.27)(100,85.85)};
\addlegendentry{SAFC, variable}

\end{axis}
\end{tikzpicture}
\caption{Bernulli arrival}
\end{subfigure}
\begin{subfigure}[b]{0.45\textwidth}
\begin{tikzpicture}
\begin{axis}[
    width=0.95\textwidth,
    legend style={at={(0.05,0.70)},anchor=south west},
    cycle list name=black white,
    xlabel={Workload},
    ylabel={Bandwidth utilization},
    xmajorgrids,
    ymajorgrids,
    zmajorgrids,
    xminorticks=true,
    xticklabel=\pgfmathprintnumber{\tick}\,$\%$,  
    yminorticks=true,
    yticklabel=\pgfmathprintnumber{\tick}\,$\%$,
    xmin=10, xmax=100,
    ymin=10, ymax=100,
    xtick={0,20,...,100},
    ytick={0,10,20,...,100},
]

\addplot coordinates {(7.798,7.795)(16.2,16.19)(25.35,25.3)(35.37,35.36)(46.13,46.07)(58.05,57.72)(70.91,69.81)(84.52,80.98)(96.79,85.91)(99.69,87.92)};
\addlegendentry{iSLIP, fixed}

\addplot coordinates {(7.798,6.494)(16.2,13.56)(25.35,21.15)(35.37,29.44)(46.13,38.42)(58.05,47.98)(70.91,58.6)(84.52,67.24)(96.79,71.68)(99.69,73.03)};
\addlegendentry{iSLIP, variable}

\addplot coordinates {(7.798,7.795)(16.2,16.19)(25.37,25.32)(35.37,35.31)(46.17,46.14)(58.01,57.8)(70.79,69.16)(84.52,82.73)(96.99,89.13)(99.76,90.12)};
\addlegendentry{SAFC, fixed}

\addplot coordinates {(7.798,6.532)(16.2,13.58)(25.37,21.06)(35.37,29.44)(46.17,38.68)(58.01,48.25)(70.79,57.55)(84.52,69.33)(96.99,74.2)(99.76,75.43)};
\addlegendentry{SAFC, variable}

\end{axis}
\end{tikzpicture}
\caption{Bursty arrival}
\end{subfigure}
\caption{Bandwidth utilization versus workloads}
\label{fig:bandwidth}
\end{figure*}
Figure~\ref{fig:bandwidth} shows the bandwidth utilization of the RIFL network under Bernoulli arrival and bursty arrival with different workloads. As the figure shows, the bandwidth utilization for variable-size packets is slightly lower than that of fixed-size packets. The reason is that RIFL Layer 2 utilizes a fixed packet length of 256 bytes, resulting in the necessity of padding for packets whose size is not an integer multiple of 256 bytes. The lowest bandwidth utilization, at 73\%, occurs under the condition of bursty traffic with variable-size packets, specifically as the workload approaches 100\%.

For both fixed-size packets and variable-size packets, the bandwidth utilization gap between the RIFL Layer 2 switch and the SAFC switch is within 3\%.

\subsubsection{Latency}

\begin{table*}[ht]
\centering
\caption{\label{tab:latency}Latencies of RIFL Layer 2 with iSLIP scheduler versus with an ideal SAFC switch}
\resizebox{0.9\textwidth}{!}{\begin{tabular}{|c|c|c|c|c|c|c|c|c|c|}
\hline
traffic pattern & scheduler & workload & latency 1\% & latency 50\% & latency 75\% & latency 90\% & latency 95\% & latency 99\% & latency 100\% \\ \hline
Bernoulli & iSLIP & 30\%  & 18 & 18  & 18  & 19  & 20  & 21  & 30  \\ \hline
Bernoulli & SAFC    & 30\%  & 18 & 18  & 18  & 19  & 19  & 20  & 30  \\ \hline
Bernoulli & iSLIP & 60\%  & 18 & 19  & 20  & 21  & 23  & 26  & 52  \\ \hline
Bernoulli & SAFC    & 60\%  & 18 & 18  & 19  & 20  & 21  & 24  & 52  \\ \hline
Bernoulli & iSLIP & 90\%  & 18 & 25  & 32  & 43  & 54  & 84  & 348 \\ \hline
Bernoulli & SAFC    & 90\%  & 18 & 20  & 24  & 29  & 35  & 54  & 260 \\ \hline
Bernoulli & iSLIP & 100\% & 18 & 213 & 361 & 476 & 524 & 573 & 611 \\ \hline
Bernoulli & SAFC    & 100\% & 18 & 65  & 129 & 219 & 283 & 401 & 574 \\ \hline
Burst     & iSLIP & 30\%  & 18 & 19  & 21  & 26  & 30  & 42  & 105 \\ \hline
Burst     & SAFC    & 30\%  & 18 & 18  & 20  & 23  & 26  & 36  & 117 \\ \hline
Burst     & iSLIP & 60\%  & 18 & 20  & 25  & 33  & 40  & 63  & 325 \\ \hline
Burst     & SAFC    & 60\%  & 18 & 19  & 23  & 30  & 36  & 56  & 179 \\ \hline
Burst     & iSLIP & 90\%  & 18 & 52  & 96  & 161 & 211 & 323 & 683 \\ \hline
Burst     & SAFC    & 90\%  & 18 & 26  & 43  & 75  & 105 & 184 & 440 \\ \hline
Burst     & iSLIP & 100\% & 18 & 57  & 107 & 178 & 229 & 339 & 683 \\ \hline
Burst     & SAFC    & 100\% & 18 & 27  & 46  & 81  & 113 & 194 & 436 \\ \hline
\end{tabular}}
\end{table*}

Table \ref{tab:latency} presents the tail latency for the RIFL (iSLIP) switch and the SAFC switch at several percentiles across different workloads, under Bernoulli arrival and bursty arrival respectively. Note that the unit of latency in the table is cell time. For a 100Gbps RIFL switch, each cell time is 20 nanoseconds.

Under Bernoulli arrival, when the workload is no more than 90\%, the latency remains within 90 cell times for 99\% of the packets. When the workload exceeds 90\%, the latency increases significantly, but even at a 100\% workload, the 99\% tail latency does not exceed 40\% more than that of an SAFC switch.

Under bursty arrival, when the workload is no more than 60\%, the packet latency tends to be higher compared to Bernoulli arrival for the same workload. However, when the workload exceeds 60\%, the latency under bursty arrival becomes lower than Bernoulli arrival. This is due to two factors: Firstly, the overall bandwidth utilization is lower under bursty arrival, reducing congestion. Secondly, the per-VOQ flow control of the RIFL Layer 2 combined with a shallow buffer allows backpressure to be transmitted more accurately and quickly to the data sender.

In general, we found that the latency data obtained from our simulations is much lower than what is presented in iSLIP. We believe this is due to the low latency of RIFL layer 1, which allows us to equip shallow VOQ buffers in the switch, and the results of our usage of per-VOQ flow control.

\subsection{Resource Cost and Fmax}
\begin{table}[h]
\centering
\caption{\label{tab:resource}Resource usage of 3-SLIP and Batcher Banyan for different switch sizes}

\begin{tabular}{|c|c|c|c|}
\hline
               & LUTs   & FFs    & Fmax     \\ \hline
\begin{tabular}[c]{@{}c@{}}3-ISLIP\\ 16 ports\end{tabular}        & 4596   & 7938   & 529.7MHz \\ \hline
\begin{tabular}[c]{@{}c@{}}3-ISLIP\\ 32 ports\end{tabular}        & 9051   & 7510   & 412.7MHz \\ \hline
\begin{tabular}[c]{@{}c@{}}3-ISLIP\\ 64 ports\end{tabular}        & 112497   & 83406   & 214.7MHz \\ \hline
\begin{tabular}[c]{@{}c@{}}Batcher Banyan\\ 16 ports\end{tabular} & 88025 & 94125 & 231.7MHz \\ \hline 
\begin{tabular}[c]{@{}c@{}}Batcher Banyan\\ 32 ports\end{tabular} & 133442 & 268934 & 206.4MHz \\ \hline 
\end{tabular}
\end{table}

Table~\ref{tab:resource} presents the resource usages and the maximum operating frequencies (Fmax) for the 3-SLIP scheduler and 100 Gbps Batcher Banyan switch used in the RIFL switch. We implemented this design on a Xilinx Versal FPGA~\cite{Versal}. Due to the spatial complexity of the iSLIP scheduler, which is $N^2$, its Fmax decreases rapidly as the port number increases. This aligns with our expectations. Despite the challenges, the 64-port iSLIP scheduler can achieve a promising frequency of 214 MHz on an FPGA, which gives us optimism regarding its performance on an ASIC implementation.

The Batcher Banyan network with 64 ports cannot be implemented on our FPGA because it cannot be placed on a single SLR (Super Logic Region) of the FPGA. If the module is placed on two different SLRs, the design cannot be successfully routed due to insufficient routing resources between the SLRs of the FPGA. However, a 64-port batcher banyan switch should be easily implementable on an ASIC, as ASICs do not have the SLR constraints.

\section{Conclusion}
\label{SECTION:conclusion}
We have successfully designed and implemented the RIFL Layer 2, applying it to intra-rack communication. It provides low-latency, lossless communication for intra-rack connections. If this paper is accepted, all the work developed within it, apart from the PCIe DMA engine that is protected under an IP agreement, will be open-sourced by us. \section{Future Work}
\label{SECTION:futurework}
In this paper, we implemented the RIFL switch using the Batcher Banyan network combined with the iSLIP scheduler. It can effectively accomplish tasks within a rack. However, if we wish to apply the RIFL network in a broader context, a high-radix switch is required, i.e., a switch with 256 ports. We are currently in the process of developing a light-weight MMM switch architecture, a project that's still underway. The preliminary results suggests this architecture outperforms the iSLIP when paired with a distributed scheduler, meanwhile enabling it to support a smaller packet size at Layer 2.

If we aim to apply the RIFL network across an entire data center, another future task is the enhancement of congestion control. Although flow control and segmented retransmission can ensure a lossless network, congestion control is still required to guarantee quality of service.

\bibliographystyle{ieeetr}
\bibliography{bibliography}

\begin{thebibliography}{10}

\bibitem{benson2010network}
T.~Benson, A.~Akella, and D.~A. Maltz, ``{Network traffic characteristics of data centers in the wild},'' in {\em Proceedings of the 10th ACM SIGCOMM conference on Internet measurement}, pp.~267--280, 2010.

\bibitem{metcalfe1976ethernet}
R.~M. Metcalfe and D.~R. Boggs, ``{Ethernet: Distributed packet switching for local computer networks},'' {\em Communications of the ACM}, vol.~19, no.~7, pp.~395--404, 1976.

\bibitem{Infiniband}
``{Infiniband Architecture Specification}.'' \url{https://www.infinibandta.org}, 2020.

\bibitem{shen2022rifl}
Q.~Shen, J.~Zheng, and P.~Chow, ``{RIFL: a reliable link layer network protocol for data center communication},'' {\em Journal of Optical Communications and Networking}, vol.~14, no.~3, pp.~111--126, 2022.

\bibitem{shen2021rifl}
Q.~Shen, J.~Zheng, and P.~Chow, ``{RIFL: A Reliable Link Layer Network Protocol for FPGA-to-FPGA Communication},'' in {\em The 2021 ACM/SIGDA International Symposium on Field-Programmable Gate Arrays}, pp.~148--148, 2021.

\bibitem{pcie6}
D.~Das~Sharma, ``{PCI Express 6.0 Specification: A Low-Latency, High-Bandwidth, High-Reliability, and Cost-Effective Interconnect With 64.0 GT/s PAM-4 Signaling},'' {\em IEEE Micro}, vol.~41, no.~1, pp.~23--29, 2021.

\bibitem{mckeown1999islip}
N.~McKeown, ``{The iSLIP scheduling algorithm for input-queued switches},'' {\em IEEE/ACM transactions on networking}, vol.~7, no.~2, pp.~188--201, 1999.

\bibitem{dpdk}
``{Data Plane Development Kit (DPDK)}.'' \url{https://www.dpdk.org/}.

\bibitem{van2019hoti}
S.~Van~Doren, ``Hoti 2019: compute express link,'' in {\em 2019 IEEE Symposium on High-Performance Interconnects (HOTI)}, pp.~18--18, IEEE, 2019.

\bibitem{IEEE802.3}
``{IEEE Standard for Information technology-- Local and metropolitan area networks-- Specific requirements-- Part 3: CSMA/CD Access Method and Physical Layer Specifications Amendment 4: Media Access Control Parameters, Physical Layers, and Management Parameters for 40 Gb/s and 100 Gb/s Operation},'' {\em IEEE Std 802.3ba-2010 (Amendment to IEEE Standard 802.3-2008)}, pp.~1--457, 2010.

\bibitem{scott2006blackwidow}
S.~Scott, D.~Abts, J.~Kim, and W.~J. Dally, ``{The blackwidow high-radix clos network},'' {\em ACM SIGARCH Computer Architecture News}, vol.~34, no.~2, pp.~16--28, 2006.

\bibitem{dong2011non}
Z.~Dong and R.~Rojas-Cessa, ``{Non-blocking memory-memory-memory Clos-network packet switch},'' in {\em 34th IEEE Sarnoff Symposium}, pp.~1--5, IEEE, 2011.

\bibitem{chrysos2006scheduling}
N.~Chrysos and M.~Katevenis, ``{Scheduling in Non-Blocking Buffered Three-Stage Switching Fabrics.},'' in {\em INFOCOM}, vol.~6, pp.~1--13, 2006.

\bibitem{sapountzis2005benes}
G.~Sapountzis and M.~Katevenis, ``{Benes switching fabrics with O (N)-complexity internal backpressure},'' {\em IEEE Communications Magazine}, vol.~43, no.~1, pp.~88--94, 2005.

\bibitem{oki2002concurrent}
E.~Oki, Z.~Jing, R.~Rojas-Cessa, and H.~J. Chao, ``{Concurrent round-robin-based dispatching schemes for Clos-network switches},'' {\em IEEE/ACM Transactions On Networking}, vol.~10, no.~6, pp.~830--844, 2002.

\bibitem{chaney1997design}
T.~Chaney, J.~A. Fingerhut, M.~Flucke, and J.~S. Turner, ``{Design of a gigabit ATM switch},'' in {\em Proceedings of INFOCOM'97}, vol.~1, pp.~2--11, IEEE, 1997.

\bibitem{narasimha1988batcher}
M.~J. Narasimha, ``{The Batcher-Banyan self-routing network: universality and simplification},'' {\em IEEE Transactions on Communications}, vol.~36, no.~10, pp.~1175--1178, 1988.

\bibitem{portCountEffect}
``{The Effect of Switch Port Count in Clos Topology}.'' \url{https://elegantnetwork.github.io/posts/Effect-of-Switch-Port-Count/}, 2020.

\bibitem{micro}
{Correa, Roberto Sanchez and David, Jean Pierre}, ``{Ultra-low latency communication channels for FPGA-based HPC cluster},'' {\em {Integration}}, vol.~63, pp.~41--55, 2018.

\bibitem{cisco}
N.~McKeown, ``{A fast switched backplane for a gigabit switched router},'' {\em Business Communications Review}, vol.~27, no.~12, pp.~1--30, 1997.

\bibitem{plummer1982arp}
{David C. Plummer}, ``{Ethernet Address Resolution Protocol: Or converting network protocol addresses to 48.bit Ethernet address for transmission on Ethernet hardware}.'' \url{https://tools.etf.org/html/rfc826}, 1982.

\bibitem{PCIE4}
``{UltraScale+ Device Integrated Block for PCI Express}.'' \url{https://www.xilinx.com/products/intellectual-property/pcie4-ultrascale-plus.html}, 2022.

\bibitem{tamir1988high}
Y.~Tamir and G.~L. Frazier, ``High-performance multi-queue buffers for {VLSI} communications switches,'' {\em ACM SIGARCH Computer Architecture News}, vol.~16, no.~2, pp.~343--354, 1988.

\bibitem{gupta1999designing}
P.~Gupta and N.~McKeown, ``Designing and implementing a fast crossbar scheduler,'' {\em IEEE micro}, vol.~19, no.~1, pp.~20--28, 1999.

\bibitem{L1switch}
{Arista}, ``{Arista 7130 FPGA-enabled Switches}.'' \url{https://www.arista.com/assets/data/pdf/7130-FPGA-Switches.pdf}.

\bibitem{VCU128}
{Xilinx}, ``{Virtex UltraScale+ HBM VCU128 FPGA Evaluation Kit}.'' \url{https://www.xilinx.com/products/boards-and-kits/vcu128.html#resources}.

\bibitem{Versal}
{Xilinx}, ``{Versal ACAP}.'' \url{https://www.xilinx.com/products/silicon-devices/acap/versal.html#documents}.

\end{thebibliography}

\end{document}